\documentclass[12pt]{iopart}

\usepackage{subfigure}

\usepackage{graphicx}
\usepackage{mathrsfs}
\usepackage[british]{babel}

\expandafter\let\csname equation*\endcsname\relax 
\expandafter\let\csname endequation*\endcsname\relax 

\usepackage{amsbsy,amssymb,amsmath,bm}
\usepackage{graphicx,color,epsfig,rotate}

\usepackage{multirow}

\begin{document}

\title[Logarithmic observables in percolation]{Logarithmic observables in critical percolation}

\author{Romain Vasseur$^{1,2}$, Jesper Lykke Jacobsen$^{2,3}$ and Hubert Saleur$^{1,4}$}
\address{${}^1$Institut de Physique Th\'eorique, CEA Saclay,
91191 Gif Sur Yvette, France}
\address{${}^2$LPTENS, \'Ecole Normale Sup\'erieure, 24 rue Lhomond, 75231 Paris, France}
\address{${}^3$Universit\'e Pierre et Marie Curie, 4 place Jussieu, 75252 Paris, France}
\address{${}^4$Department of Physics,
University of Southern California, Los Angeles, CA 90089-0484}

\eads{\mailto{romain.vasseur@cea.fr}, 
      \mailto{jesper.jacobsen@ens.fr},
      \mailto{hubert.saleur@cea.fr}}

\date{\today}

\begin{abstract}

Although it has long been known that the proper quantum field theory 
description of critical percolation involves a  logarithmic conformal
field theory (LCFT), no direct consequence of this has been observed so far. 
Representing critical bond percolation as the $Q \to 1$ limit of the $Q$-state Potts model, and
analyzing the underlying $S_Q$ symmetry of the Potts spins, we identify a class of simple observables whose two-point
functions scale logarithmically
for $Q \to 1$. The logarithm originates from the mixing of the energy operator with a logarithmic partner that we identify as the
field that creates two propagating clusters. In $d=2$ dimensions this agrees with general LCFT results, and in particular the
universal prefactor of the logarithm can be computed exactly. We confirm its numerical value by extensive Monte-Carlo simulations.

\end{abstract}



\newpage

\section{Introduction} 

The analysis of two-dimensional critical geometrical problems such as percolation or the self-avoiding
walks (SAW) in terms of conformal field theory (CFT) involves
features more complicated than those appearing in minimal (e.g., Ising or $3$-state Potts) models.
One might think that these complications are due to the non-local nature of the interesting geometric observables, such as
connectivities of clusters and loops. This viewpoint is however somewhat misleading, since genuine non-locality would 
break the invariance under conformal transformations. In fact the above problems definitely {\em are} conformally invariant
and their non-locality is only apparent: It can be traded for non-unitarity, by reformulating
these models in terms of vertex models or supersymmetric spin chains. The origin of the difficulty in building a CFT description
of percolation or SAWs is precisely this non-unitarity, which turns out to have deeper consequences than that of certain minimal models, such as the Lee-Yang edge singularity.

One of these consequences is  that the operator algebra underlying the lattice description---or the Virasoro algebra that emerges
in the continuum limit---of percolation or SAW problems involves representations that are not fully reducible. This leads to
indecomposability and, ultimately, to the appearance of logarithms in correlation functions. This feature follows from the fact that
the models cited above have central charge $c=0$, and hence it is necessary to go `outside'
their minimal Kac table (made of just the identity operator) to describe non-trivial physical observables.
This extension leads to {\sl logarithmic CFTs} (LCFTs) whose study has attracted considerable interest over the last few years,
as their potential role in condensed matter as well as string theory applications has become more evident.

The best known property of LCFTs at $c=0$ is the existence of a logarithmic partner to the stress-energy tensor, whose presence
is necessary (under some circumstances at least) to avoid the `$c=0$ catastrophe' \cite{Gurarie93}. Associated with this partner is a
universal number, the so-called `$b$ number' or logarithmic coupling. It was suggested that $b$ is  a sort of effective central charge
\cite{Gurarie99} that could potentially be used to distinguish different $c=0$ theories \cite{GurarieLudwig}. While there has been a lot
of work on the determination of the $b$ number and its relation with abstract module properties of the Virasoro algebra
\cite{MathieuRidout}, it is only very recently that methods have been devised to measure it in a numerical experiment. These methods 
\cite{Dubail,Vasseur1} are however rather indirect---in particular, they require a proper quantization scheme---and thus
to this day no simple physical procedure to  determine $b$ in a real experiment has been proposed.

The indecomposability that leads to the existence of a logarithmic partner to the stress-energy tensor can be traced back to the
peculiarities of the $Q \to 1$ (resp.\ $n \to 0$) limit that relate percolation (resp.\ SAW) to the $Q$-state Potts
(resp.\ $\mathcal{O}(n)$ spin) model. It is then possible to predict other striking consequences~\cite{Cardylog, Rahimi, Rahimi1}  that should be observed at $c=0$, such as logarithmic terms in certain correlation functions. These logarithms
arise from degeneracies, where several operators that are distinguishable at $Q$ (resp. $n)$ generic `mix' in the limit $Q\to 1$
(resp.\ $n\to 0$). This mixing is the physical phenomenon resulting from the indecomposability in the formal algebraic
description. Nevertheless, a clear physical interpretation of such logarithmic correlators is often difficult to find, 
and also hard to access numerically since the logarithm multiplies a much stronger power law singularity. 
Direct consequences of the logarithmic nature of the underlying
CFT have thus, to this day, remained unobserved. 

The purpose of this Letter is to present a new example of a physical observable in the $Q$-state Potts model which, one the one hand,
exhibits a {\em pure} logarithmic dependency (with no multiplying power law) in the $Q \to 1$ percolation limit, whose universal
prefactor we compute analytically in $d=2$ dimensions, and, on the other hand, has a sufficiently simple geometric formulation in terms of percolation clusters to make possible a numerical study. And indeed our thorough Monte Carlo simulations nicely confirm both the logarithmic
scaling and our result for the universal prefactor.

\section{Percolation, Potts model, and continuum limit} 

We first recall the well-known reformulation of bond percolation
as the $Q \rightarrow 1$ limit of the Potts model.
The partition function of the Potts model reads
\begin{equation}
 Z = \sum_{\sigma} \prod_{(ij) \in E}
 \exp \left( K \delta_{\sigma_i,\sigma_j} \right) \,,
 \label{Potts}
\end{equation}
where $K$ is the coupling between spins $\sigma_i = 1,2,\ldots,Q$
along the edges $E$ of some lattice $G$. 
The Kronecker symbol $\delta_{\sigma_i,\sigma_j}$ equals 1 if $\sigma_i =
\sigma_j$, and 0 otherwise. Universal properties depend only on the dimension
$d$, and not on the precise choice of $G$. Although the main conclusions  
of this Letter should be valid in any dimension, for the sake of simplicity, we shall 
restrict ourselves to $d=2$ in the following. 
The generalization of our results in higher dimensions
will be discussed in the conclusion.
We therefore suppose that $G$ is the square lattice.
The transition line---which gives rise to a critical
theory for $0 \le Q \le 4$---is then given by the selfduality
criterion ${\rm e}^K = 1 + \sqrt{Q}$.

We can expand $Z$ by rewriting the local Boltzmann 
weight as $\exp \left( K \delta_{\sigma_i,\sigma_j} \right) = 1 + \frac{p}{1-p} \delta_{\sigma_i,\sigma_j}$, 
with $p = 1-{\rm e}^{-K}$. The set of edges $(ij) \in E$ for which we have the term $\frac{p}{1-p}
\delta_{\sigma_i,\sigma_j}$ are called `occupied bonds'. 
They define a graph $H$, whose
connected components are known as Fortuin-Kasteleyn (FK)
clusters~\cite{FK72}. Since spins belonging to the same FK cluster are
aligned, we can perform the sum over $\{\sigma_i\}$ in~\eqref{Potts} 
to obtain
\begin{equation}
 Z \propto \sum_{H} Q^{{\rm \sharp clusters}(H)} p^{{\rm \sharp edges}(H)} (1-p)^{\left|E \right| - {\rm \sharp edges}(H)}\,.
 \label{PottsFK}
\end{equation}
In this formulation, the number of colors $Q$ can be thought of as 
a real parameter, and the limit $Q \rightarrow 1$ yields a sum over
bond percolation configurations, with a probability $p$ per occupied bond.
This model is critical for $p=p_c=\frac{1}{2}$, and although the partition
function $Z=1$ is trivial, the correlation functions of the model
capture the salient geometrical properties of critical percolation clusters.

The continuum limit of the critical Potts model is described by a CFT.
To establish this standard result, one first notices that the (outer and inner) hulls
of the FK clusters constitutes a gas of loops.
This in turn defines a height model (the loops being contour lines of the height)
which can be argued to renormalize towards a Coulomb Gas (CG), 
that is, a compactified free boson with Lagrangian
density $\mathcal{L}=\frac{g}{2\pi} (\nabla \phi)^2$, along with additional
`electric charges' at infinity. The stiffness $g \in \left[2,4 \right]$
is given by $Q=2+2 \cos \frac{\pi g}{2}$, so that percolation has $g=\frac{8}{3}$.
Over the past thirty years, this mapping has allowed physicists to compute 
many interesting geometrical properties for percolation (and for the Potts model in general), 
including crossing probabilities and critical exponents. More recently, the continuum limit
has been studied by mathematicians under the name ${\rm SLE}_\kappa$, where
$\kappa = 16/g$.

\section{Symmetric group $S_{Q}$ and operators in the Potts model} 
\label{sec:reptheory}

To understand how logarithms appear in percolation,
the key idea is the study of the symmetric
group $S_Q$ in the (formal) limit $Q \to 1$ (see also~\cite{Cardylog}). We expect scaling fields
of the underlying CFT to transform as irreducible representations (irreps) under the 
$S_Q$ symmetry of the Potts model.

Consider first observables
acting on a single spin $\mathcal{O}(\sigma_i)$, for $Q$ integer. 
Obviously, any such operator can be decomposed 
onto a basis of $Q$ generators as $\mathcal{O}(\sigma_i) = \sum_{a=1}^{Q} c_a \delta_{\sigma_i,a}$.
The action of $S_Q$ defines a representation
of dimension $Q$, which however is reducible.
Indeed, $\delta_{\sigma_i,a}$ can be decomposed onto an invariant $1 = \sum_{a=1}^{Q} \delta_{\sigma_i,a}$, 
and an irrep $\varphi_a = \delta_{\sigma_i,a} - 1/Q$ with $\sum_{a=1}^Q \varphi_a=0$
of dimension $Q-1$. Obviously $1$ corresponds to the identity operator,
while $\varphi_a$ is the magnetization, or order parameter operator, of the Potts model.
The two-point
function $\langle \varphi_a(\sigma_i) \varphi_b(\sigma_j) \rangle$ vanishes if $i$ and $j$
belong to different FK clusters (the sums over $\sigma_i$ and $\sigma_j$ being independent).
At the critical point, $\langle \varphi_a(\sigma_i) \varphi_b(\sigma_j) \rangle$ 
decays algebraically as $r^{-2 \Delta_{\varphi}}$, where the
(bulk) critical exponent $\Delta_{\varphi}=\frac{(6-g)(g-2)}{8g}$ can be 
computed within the CG setup~\cite{MagneticPotts,NienhuisLoop}.

Nothing particular happens when one takes $Q \to 1$
in these expressions. However, the limit $Q \rightarrow 0$
in $\langle \varphi_a(\sigma_i) \varphi_b(\sigma_j) \rangle$  is ill-defined, and although we will not discuss it in details here,
this is actually responsible for the occurrence 
of logarithms at the level of the identity operator. The $Q \to 0$ field theory of
free (symplectic) fermions can be interpreted
geometrically in terms of dense polymers, or spanning trees on $G$, and in this 
latter context several
similar results were obtained using exact combinatorial methods~\cite{Ivashkevich,SpannLog1,SpannLog2}.
This theory is also related to the logarithmic form
of the Gaussian propagator in $d=2$ and to the asymptotic behavior of the 
equivalent resistance in an infinite network of resistors {\it via} the Kirchhoff theorem~\cite{Wu}.

In the remainder of the Letter we focus on the $Q \to 1$ percolation case for which one needs to
consider observables acting on {\em two} nearest-neighbors spins
$\mathcal{O}(r_i)\equiv\mathcal{O}(\sigma_i, \sigma_{i+1})$ in order to
recover logarithms. We impose the constraint $\sigma_i \neq \sigma_{i+1}$, whence
these observables are $Q \times Q$ (symmetric) matrices
with zero elements on the diagonal.
As before, the starting point is the basis elements $\delta_{\sigma_{i},a}\delta_{\sigma_{i+1},b}$
which should be symmetrized in $\{\sigma_i,\sigma_{i+1}\}$ in order to obtain symmetric matrices.
The decomposition of this representation of $S_Q$ is straightforward and we find 
\begin{equation}
\underbrace{\left( \dfrac{Q(Q-1)}{2} \right)}_{\substack{{\rm Symmetric \ matrices} \ \mathcal{O}(\sigma_i, \sigma_{i+1}) \\ \rm with \ zero \ diagonal }} = \left( 1 \right) \oplus \left( Q-1 \right)  \oplus \underbrace{\left(\dfrac{Q(Q-3)}{2} \right)}_{\sum_{\sigma_i}\mathcal{O}(\sigma_i, \sigma_{i+1})=0} \,,
\label{eqirrepdims}
\end{equation}  
where we denoted the representations by their dimensions.
The explicit expression for the generators of these representations reads
\begin{subequations}
\label{eqPott2Spin}
\begin{eqnarray}
\fl E(\sigma_i,\sigma_{i+1}) &=& \delta_{\sigma_i \neq \sigma_{i+1}}  \,, \\
\fl \phi_a(\sigma_i,\sigma_{i+1}) &=&  \delta_{\sigma_i \neq \sigma_{i+1}} \left( \varphi_a(\sigma_i)+\varphi_a(\sigma_{i+1}) \right)   \,, \\
\fl \hat{\psi}_{ab} (\sigma_i,\sigma_{i+1})&=& \delta_{a \neq b} \left( \delta_{\sigma_i, a} \delta_{\sigma_{i+1}, b} +\delta_{\sigma_i, b} \delta_{\sigma_{i+1}, a} -\frac{1}{Q-2}  \left(\phi_a+
\phi_b \right)  -\frac{2}{Q(Q-1)} E \right) \,, \label{hatpsi}
\end{eqnarray} 
\end{subequations}
where we recall that $\varphi_a(\sigma_i) = \delta_{\sigma_i,a} - 1/Q$. The `scalar' $E$ is obviously a one-dimensional irrep.
The operator $\phi_a$ satisfies $\sum_a \phi_a = 0$, and transforms like the `vector' $\varphi_a$.
The `tensor' $\hat{\psi}_{ab} = \hat{\psi}_{ba}$ satisfies $\sum_{a (\neq b)} \hat{\psi}_{ab} = 0$ for any $b$.
This last irrep thus has $Q(Q-1)/2-Q=Q(Q-3)/2$ independent elements, as expected from (\ref{eqirrepdims}).

The fact that we obtained three independent operators acting on two spins can be 
understood physically in terms of fusion of one-spin operators (see also \cite{CardyIncipient}). As $\sigma_i \neq \sigma_j$, all three
operators act on two distinct FK clusters. The operator with highest symmetry, $\hat{\psi}$, will
then impose that the two clusters propagate until they encounter another $\hat{\psi}$ operator.
In section~\ref{sec:geom} we further develop this geometric interpretation.

\section{Percolation limit and Jordan cell for the energy operator} 
\label{sec:log}

We  can already see from \eqref{eqPott2Spin} that $\hat{\psi}_{ab} $ becomes ill-defined 
in the  (formal) $Q\rightarrow 1$ limit. We shall now see that a well-defined limit is obtained by mixing
$\hat{\psi}_{ab}$ with $E$.
Let us first study these operators from a quantum field theory point of view.
The energy operator is given by 
$\varepsilon (r_i) \equiv E(\sigma_i,\sigma_{i+1}) - \langle E \rangle$,
where we subtracted the bulk expectation value of $E$ so as to obtain a well-defined scaling field.
This subtraction does not change the representation theoretic considerations, so $\varepsilon(r_i)$ remains an irrep.
This field corresponds to the thermal perturbation of the Potts model, and its bulk scaling
dimensions is $\Delta_{\varepsilon} = \frac{6}{g}-1 $~\cite{ThermalPotts,NienhuisLoop}. 
In the CFT with generic real $Q$, we expect the following form for the two point function
\begin{equation}
\displaystyle \langle \varepsilon (r)  \varepsilon (0) \rangle = \tilde{A}(Q) (Q-1) r^{-2 \Delta_{\varepsilon}(Q)} \,,
\end{equation} 
where $\tilde{A}(Q)$ is a regular function of $Q$, with a finite non-zero limit $\tilde{A}(1)$ for $Q \to 1$.
The reasons why $\langle \varepsilon (r)  \varepsilon (0) \rangle$ should vanish at $Q=1$ is very natural from
a lattice point of view, and was recently argued in the context of bulk logarithmic CFT \cite{Vasseur2}.

The above fusion considerations imply that the `tensor' operator $\hat{\psi}_{ab}$ can be identified, 
with the so-called 4-leg watermelon operator \cite{NienhuisLoop} (the two propagating clusters
correspond to four propagating hulls).
Its bulk scaling dimension $\Delta_{\hat{\psi}} = \frac{(4+g)(3g-4)}{8g}$ follows from a CG computation
\cite{SaleurDuplantier87}.
We thus deduce the form of the two-point function
\begin{multline}
\label{eqTwoPointpsiHat}
\langle \hat{\psi}_{ab} (r) \hat{\psi}_{cd} (0) \rangle = \frac{2 A(Q)}{Q^2} \left( \delta_{ac} \delta_{bd} + \delta_{ad} \delta_{bc} - \frac{1}{Q-2} \left(\delta_{ac}+\delta_{ad}+\delta_{bc}+\delta_{bd} \right)  \right. \\
+\left. \frac{2}{(Q-1)(Q-2)} \right)
 \times r^{-2 \Delta_{\hat{\psi}}(Q)},
\end{multline} 
where $A(Q)$ is again a regular function of $Q$ when $Q\rightarrow 1$, and the factor $2/Q^2$ is purely conventional.
However, the Kronecker symbols combination is completely fixed by (\ref{hatpsi}).

In the formal limit $Q\rightarrow 1$ the two-point function~(\ref{eqTwoPointpsiHat}) diverges.
To cure this, we introduce a new field
\begin{equation}
 \tilde{\psi}_{ab}  (r) = \hat{\psi}_{ab}  (r) + \frac{2}{Q(Q-1)}  \varepsilon (r), \ a \neq b.
 \label{eq_tilde_psi}
\end{equation}
Its two-point function is easily computed and
in order to have a finite $Q \to 1$ limit, we must require $A(1)=\tilde{A}(1)$, and that $\Delta_{\varepsilon}=\Delta_{\hat{\psi}}$
at $Q=1$. The latter condition is indeed satisfied, since for $g=\frac{8}{3}$ the CG results read
$ \Delta_{\varepsilon}=\Delta_{\hat{\psi}}= \frac{5}{4}$.
Assuming also the former, the two-point function of $\tilde{\psi}_{ab}$ has a well-defined $Q \to 1$ limit:
\begin{equation}
\label{eq_PercoLog}
 \langle \tilde{\psi}_{ab}  (r) \tilde{\psi}_{cd}  (0) \rangle = 2 A(1) r^{- 5/2} \left[
 \left( \delta_{ac}+\delta_{ad}+\delta_{bc}+\delta_{bd} + \delta_{ac} \delta_{bd} + \delta_{ad} \delta_{bc} \right) +
 \frac{4 \sqrt{3}}{\pi} \log r \right] \,,
\end{equation} 
where we have used
\begin{equation}
\label{eqLimitbeta}
\lim_{Q \rightarrow 1} \frac{\Delta_{\hat{\psi}} - \Delta_{\varepsilon}}{Q-1} = \frac{\sqrt{3}}{\pi} \,.
\end{equation} 
We have thus identified a logarithmic two-point function (\ref{eq_PercoLog}) as a result of the
mixing (\ref{eq_tilde_psi}) of the energy operator $\varepsilon$
and the 4-leg operator $\hat{\psi}$ of the Potts model. This mixing is consistent with recent algebraic
results in bulk LCFTs~\cite{Vasseur2,GRSV}, but was obtained here from
very simple physical arguments based on the ``$S_{Q=1}$ symmetry'' of the theory. 

Logarithms in LCFTs can be associated with the non-diagonalizability of the scale 
transformation generator---{\it i.e.}, the Hamiltonian in the usual radial quantization---of the theory.
To show that this logarithm corresponds to a Jordan cell, we now study the change of the field $\tilde{\psi}_{ab}$
under a scale transformation $r \to \Lambda r$.
Recall \cite{NienhuisLoop} that $\varepsilon$ and  $\hat{\psi}_{ab}$ are both primary operators for generic $Q$
and thus transform as
\begin{subequations}
\begin{eqnarray}
\varepsilon(\Lambda r) &=& \Lambda^{-\Delta_{\varepsilon}} \varepsilon(r), \\
\hat{\psi}_{ab}(\Lambda r) &=& \Lambda^{-\Delta_{\hat{\psi}}} \hat{\psi}_{ab}(r) \,.
\end{eqnarray}
\end{subequations}
Using this, one can readily show that at $Q=1$
 \begin{equation}
\tilde{\psi}_{ab}(\Lambda r) = \Lambda^{-5/4} \left(\tilde{\psi}_{ab}(r) + \frac{2 \sqrt{3}}{\pi} \log \Lambda \ \varepsilon(r) \right).
\end{equation} 
The field $\tilde{\psi}_{ab}$ is therefore mixed with the energy operator $\varepsilon(r)$ after a scale transformation.
We stress that the appearance of logarithms 
in the two-point function (\ref{eq_PercoLog}) of the field $\tilde{\psi}$ is fully compatible with scale and conformal
invariance of the critical theory. The point is that some operators are not pure scaling operators, and get mixed
with others under a scale transformation. In other words, the scale transformation generator (or Hamiltonian)
is  non-diagonalizable, with a rank-2 Jordan cell mixing the two fields $\tilde{\psi}_{ab}$ and $\varepsilon$. 

\begin{figure}[t]
\begin{center}
\includegraphics[width=12cm]{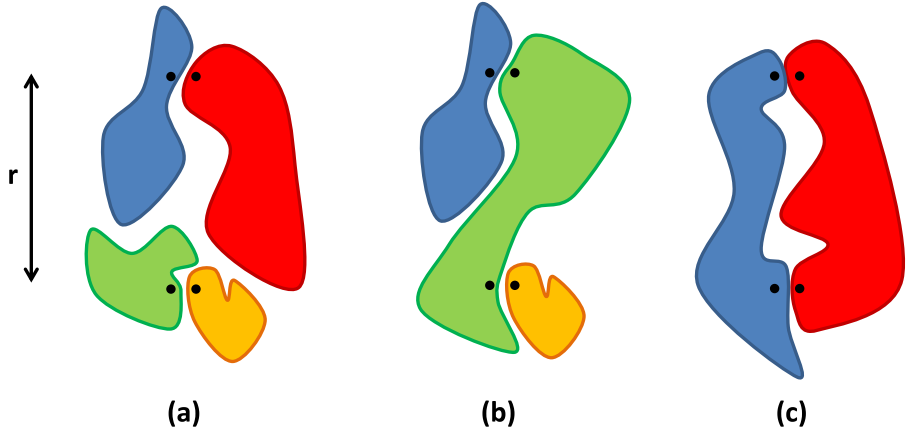}
\end{center}
\caption{Percolation configurations contributing to (a) ${\mathbb P}_0(r)$, (b) ${\mathbb P}_1(r)$ (one cluster propagating a distance $r=|r_1-r_2|$), (c) ${\mathbb P}_2(r)$ (two propagating clusters). \label{fig1configs}}
\end{figure}

\section{Geometrical interpretation}
\label{sec:geom}

We now show that (\ref{eq_PercoLog}) has a very nice physical meaning
in terms of geometrical observables in the percolation problem.
Let $i_1$ and $i_1+1$ be two nearest-neighbor points, separated from two other nearest-neighbor points 
$i_2$ and $i_2+1$ by a distance $r=|r_1-r_2|$. To be consistent with the bulk calculation of section~\ref{sec:log}, 
we consider percolation defined on a torus ({\it i.e.}, with doubly periodic boundary conditions) and we assume
that $r$ is much smaller than either period.
We define ${\mathbb P}_{\neq}$ as the (position independent) probability that two neighboring points belong to
different FK clusters. Let ${\mathbb P}_0(r)$  be the probability that $\{i_1,i_1+1,i_2,i_2+1\}$ belong to four
different FK clusters. Note that ${\mathbb P}_0(r)$ is an {\it increasing} function of $r$, with limit
$({\mathbb P}_{\neq})^2$ for $r \to \infty$. Further let ${\mathbb P}_1(r)$ be the probability that
$\{i_1,i_1+1,i_2,i_2+1\}$ belong to three different FK clusters, of which {\em one} `propagating' cluster contains
one point from $\{i_1,i_1+1\}$ and one point from $\{i_2,i_2+1\}$---there are four ways to make this choice. 
Finally, let ${\mathbb P}_2(r)$ be the probability that $\{i_1,i_1+1,i_2,i_2+1\}$ belong to two different FK clusters,
i.e., with {\em two} distinct clusters each containing one point from $\{i_1,i_1+1\}$ and one point from
$\{i_2,i_2+1\}$---this can be done in two ways. Configurations contributing to these probabilities 
are shown schematically in Fig.~\ref{fig1configs}.

Using these probabilities, one can interpret geometrically the Potts operators previously introduced.
For example, it is straightforward to show that
$\langle E \rangle = \frac{Q-1}{Q} {\mathbb P}_{\neq} $.
Recall that the energy operator $\varepsilon$ was constructed by subtracting this expectation value
from the operator $E$.
A careful analysis shows that the two-point correlator of $\hat{\psi}_{ab}$ is 
proportional to ${\mathbb P}_2(r)$ 
%
\begin{multline}
\langle \hat{\psi}_{ab} (\sigma_{i_1},\sigma_{i_1+1}) \hat{\psi}_{cd} (\sigma_{i_2},\sigma_{i_2+1}) \rangle = \frac{2}{Q^2} \left( \delta_{ac} \delta_{bd} + \delta_{ad} \delta_{bc} - \frac{1}{Q-2} \left(\delta_{ac}+\delta_{ad}+\delta_{bc}+\delta_{bd} \right)  \right. \\ 
+\left. \frac{2}{(Q-1)(Q-2)} \right)
 \times 
{\mathbb P}_2(r) \,.
\end{multline} 
We infer from \eqref{eqTwoPointpsiHat} that ${\mathbb P}_2(r) \sim A(Q) r^{-2 \Delta_{\hat{\psi}}(Q)}$.
Other correlation functions follow from the lattice description in the same way.
A direct computation shows that $\langle \hat{\psi}_{ab} \rangle = \langle \phi_a \rangle = 0$.
This also follows from the representation theory of $S_Q$, which moreover
implies the vanishing of the 	`crossed' correlation functions:
$\langle \varepsilon \hat{\psi}_{ab} \rangle = \langle \varepsilon \phi_a \rangle =
\langle \hat{\psi}_{ab} \phi_c \rangle  = 0$. The vanishing of one-point functions and crossed two-point
functions is consistent with fundamental CFT results.

To analyze the $Q \rightarrow 1$ limit from a CFT perspective, we studied the correlation functions
of the field $\tilde{\psi}_{ab}$ defined in (\ref{eq_tilde_psi}). One can repeat the very same steps
from the lattice perspective. It is convenient to write
$\tilde{\psi}_{ab}  (r_i) \equiv \hat{\psi}_{ab}  (\sigma_i,\sigma_{i+1}) + \frac{2}{Q(Q-1)}  \varepsilon (\sigma_i,\sigma_{i+1}) =  \psi_{ab}  (\sigma_i,\sigma_{i+1}) -  \langle \psi \rangle $,
where we have introduced 
\begin{equation}
 \psi_{ab}  (\sigma_i,\sigma_{i+1}) = \delta_{\sigma_i, a} \delta_{\sigma_{i+1}, b} +\delta_{\sigma_i, b}
  \delta_{\sigma_{i+1}, a} -\frac{1}{Q-2}  \left(\phi_a(\sigma_i,\sigma_{i+1})+
\phi_b(\sigma_i,\sigma_{i+1})  \right)  .
\end{equation} 
Note that $\psi_{ab}$ is not a scaling field
(since $\langle \psi \rangle = \frac{2}{Q^2} {\mathbb P}_{\neq} \neq 0$), whereas $ \tilde{\psi}_{ab}$ is.
One can show that the two-point function of $\psi_{ab}$ reads
\begin{multline}
\displaystyle \langle \psi_{ab} (r_1) \psi_{cd} (r_2) \rangle =  \frac{4}{Q^4} \left({\mathbb P}_0(r)+
 {\mathbb P}_1(r) \right)+{\mathbb P}_2(r) \\
 \times \frac{1}{Q^2} \left[ \frac{8}{Q(Q-2)} + \frac{2}{2-Q} 
 (\delta_{ac}+\delta_{ad}+\delta_{bc}+\delta_{bd}) + 2 ( \delta_{ac} \delta_{bd} + \delta_{ad} \delta_{bc} ) \right] \,,
\end{multline}  
whereas that of the corresponding scaling field $\tilde{\psi}_{ab}$ is, in the limit $Q=1$, 
\begin{multline}
 \displaystyle \langle \tilde{\psi}_{ab} (r_1) \tilde{\psi}_{cd} (r_2) \rangle =  
 2 \left( \delta_{ac}+\delta_{ad}+\delta_{bc}+\delta_{bd} + \delta_{ac} \delta_{bd} +
 \delta_{ad} \delta_{bc} \right) {\mathbb P}_2(r) \\
 + 4 \left[{\mathbb P}_0(r) + {\mathbb P}_1(r) - 2{\mathbb P}_2(r)  - {\mathbb P}_{\neq}^2 \right]  .
\end{multline}  
Comparing with \eqref{eq_PercoLog} we deduce that ${\mathbb P}_2(r) \sim A(1) r^{-5/2}$, as was
of course expected from its relation to the 4-leg operator. Meanwhile, the logarithmic term in
\eqref{eq_PercoLog} can be identified with
\begin{equation}
 \displaystyle {\mathbb P}_0(r) + {\mathbb P}_1(r) - {\mathbb P}_{\neq}^2 \sim
  A(1) \left( \theta + \frac{2 \sqrt{3}}{\pi} \log r \right) r^{-5/2} \,,
\end{equation}  
where we have added a subdominant non-universal (i.e., lattice dependent) term $\theta$.
Finally, the following combination
\begin{equation}
\label{eqLogGeomPerco}
 \displaystyle F(r) \equiv \dfrac{{\mathbb P}_0(r) + {\mathbb P}_1(r) - {\mathbb P}_{\neq}^2}
 {{\mathbb P}_2(r)} \sim \theta + \frac{2 \sqrt{3}}{\pi} \log r,
\end{equation}
cancels out the dominant power law ($r^{-5/2}$), leaving a {\rm pure} logarithmic scaling which
should be observable in numerical simulations (see below). The number
$\frac{2\sqrt{3}}{\pi} \simeq 1.1026$ is a universal constant that can be traced back
to \eqref{eqLimitbeta}.
Although the combination \eqref{eqLogGeomPerco} may look slightly complicated,
it is important to keep in mind that the logarithmic term we are after resides in
the {\em disconnected} part ${\mathbb P}_0(r)$---a similar observation holds true for
LCFTs with other values of $Q$, such as $Q \to 0$.

\begin{figure}[t]
\begin{center}
\includegraphics[width=12cm]{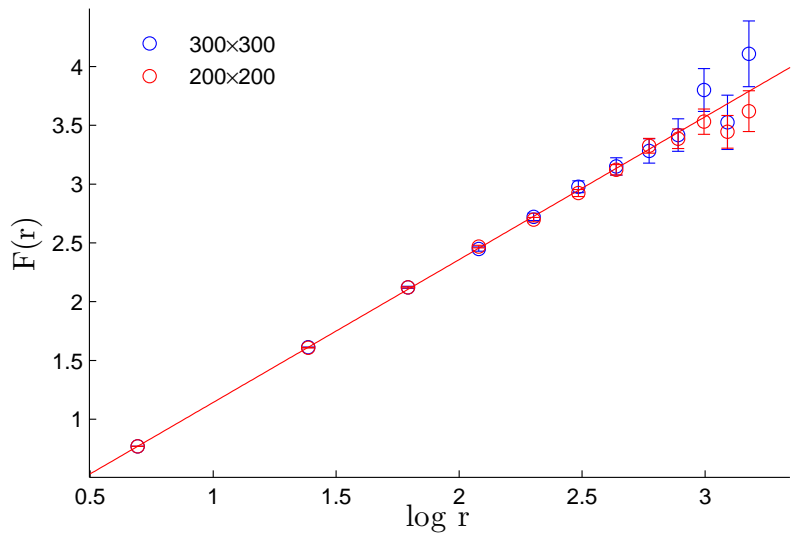}
\end{center}
\caption{Monte Carlo estimation of the function $F(r)$ defined in eq.~\eqref{eqLogGeomPerco}. Results are shown for  $200 \times 200$ and $300 \times 300$ square lattices, with no significant difference. \label{fig2}}
\end{figure}

We have checked the validity of (\ref{eqLogGeomPerco}) by performing extensive Monte Carlo
simulations on square lattices of various sizes ranging
from $150 \times 150$ to $300 \times 300$, with doubly periodic boundary conditions.
We checked that different pseudo-random number generators---including a
Mersenne-Twister algorithm~\cite{Mersenne-Twister}---led to consistent results.
Statistics were obtained on $\sim 10^3$ independent runs of $10^7$ percolation configurations
each. Results are shown in Fig.~\ref{fig2}, and are in good agreement with 
(\ref{eqLogGeomPerco}).
Careful extrapolations removing successively the first few short-distance points yields a slope $1.15 \pm 0.05$ in
good agreement with our prediction $\frac{2\sqrt{3}}{\pi} \simeq 1.1026$.

Note that although all the calculations of this Letter were made 
in the bulk, the derivation in the boundary case presents only minor differences.
In this case the scaling dimension of the energy operator should be replaced by
$\Delta_{\varepsilon}=2$, since $\varepsilon$ becomes degenerate with the stress-energy
tensor $T$ \cite{Cardy84}. Similarly, the $4-$leg watermelon exponent should be changed
\cite{DuplantierSaleurNPB87} to $\Delta_{\hat{\psi}}=\frac{3 g}{2}-2$, and
the operator $\tilde{\psi}_{ab}$ would be proportional to the well-known 
logarithmic partner $t(z)$ of the stress-energy tensor introduced 
by Gurarie~\cite{Gurarie99} and Gurarie and Ludwig~\cite{GurarieLudwig,GurarieLudwig2} 
in their  work on CFTs with central charge $c=0$.
However, it is easy to see that the limit~(\ref{eqLimitbeta}) remains unchanged 
so we expect (\ref{eqLogGeomPerco}) to hold true also if the points 
lie at a boundary. It would be interesting 
to check this numerically as well.  

\section{Conclusion}

We have found a simple geometrical observable in percolation that provides a lattice
version of a logarithmic two-point function in a $c=0$ (L)CFT. The method we used is very
similar to what was done by Cardy for disordered systems and for the $\mathcal{O}(n)$ model~\cite{Cardylog},
and is far more general in that respect than the specific logarithmic solutions \cite{Ivashkevich,SpannLog1,SpannLog2}
found in the case 
of the free-fermion dense polymers case ($Q=0$). Meanwhile, our result (\ref{eqLogGeomPerco}) for percolation allows for a direct
numerical verification, and the prefactor in front of the logarithmic term turns out to be universal.
It should be noticed that this coefficient is actually closely related to what is known as logarithmic coupling
or indecomposability parameter in the context of LCFT~\cite{GurarieLudwig,GurarieLudwig2,MathieuRidout, Vasseur1}.
We also note that, remarkably, in all the other examples we have studied, 
logarithmic terms tend to appear in {\it disconnected} observables such as ${\mathbb P}_0(r)$.
This is similar to results for disordered systems \cite{Cardylog}, and we will get back to this issue elsewhere.

The main parts of our derivation---notably the representation theory of section~\ref{sec:reptheory}---are not restricted
to the $d=2$ dimensional case. Indeed, we expect $\Delta_{\hat{\psi}} = \Delta_\varepsilon$ at $Q=1$ also for $d>2$\footnote{
This implies that the fractal dimension $d_{\rm RB}$ of the so-called ``red bonds''  (also called ``cutting bonds'')
 is related to the thermal exponent $\nu$ 
{\it via} $d_{\rm RB}=\nu^{-1}$. This is indeed a well-known percolation result~\cite{Coniglio}.},
and only the derivative (\ref{eqLimitbeta}) will change. Accordingly we expect (\ref{eqLogGeomPerco})
to remain correct for $d>2$, albeit with a different universal prefactor that could be, in principle, 
computed in a $\epsilon=6-d$ expansion.

Our results are not restricted to percolation. In particular, studying the $Q \rightarrow 2$ limit 
should also yield logarithmic observables, involving geometrical properties of Ising spin clusters.
Higher-rank correlation functions and observables acting on more spins---including the generalization of $2n$-leg watermelon
operators to $d>2$---can be worked out along the same lines. There are indications that matching these more physical observations
with formal algebraic developments should lead to further progress in our understanding of LCFTs.

In conclusion, it is important  to stress that logarithmic terms such as those we have identified would not be present for generic $Q$,
and occur solely because of the special degeneracies present at $Q=1$. This is of course quite different from logarithmic dependencies
in other non-local quantities---see e.g. \cite{JacobsenSaleur, CardyHall}---which are obtained as
derivatives of correlation functions with respect to the Boltzmann weights (such as $Q$).

\section*{Acknowledgments}

We gratefully acknowledge stimulating discussions with A.\ Gainutdinov and J.\ Cardy.
We also thank the Institut Henri
Poincar\'e for hospitality during the recent trimestre ``Advanced Conformal Field Theory and Applications''.
This work was supported by the Agence Nationale de la Recherche (grant
ANR-10-BLAN-0414: DIME).

\end{document}